\documentstyle[aps,prl,floats,epsf]{revtex}

\begin{document}

%************************************************************
%***** Title ************************************************
%************************************************************

\twocolumn[\hsize\textwidth\columnwidth\hsize\csname@twocolumnfalse%
\endcsname
\title{Local spin polarization in underdoped cuprates with
impurities}
\author{
G.\ Khaliullin$^{\text{a,b}}$,
R.\ Kilian$^{\text{a}}$,
S.\ Krivenko$^{\text{a,b}}$,
P.\ Fulde$^{\text{a}}$}
\address{$^{\text{a}}$ Max-Planck-Institut f\"ur Physik komplexer 
Systeme, N\"othnitzer Strasse 38, D-01187 Dresden, Germany}
\address{$^{\text{b}}$ Kazan Physicotechnical Institute of the Russian 
Academy of Sciences, 420029 Kazan, Russia}
\date{24 September 1998}
\maketitle

%************************************************************
%***** Abstract *********************************************
%************************************************************

\begin{abstract}
We present a theory of magnetic (Ni) and nonmagnetic (Zn) 
impurities substituted into planar Cu sites in the normal state of
underdoped cuprates exhibiting a spin gap. Both types of impurities induce
magnetic moments on neighboring Cu sites. In the case of Ni these
moments partially screen the inherent impurity spin,
resulting in an effective $S=1/2$ moment. The characteristic
Kondo scale is found to have a power-law dependence on the 
coupling constant. We investigate the spatial shape of the 
impurity-induced spin density, taking into account the 
presence of short-ranged AF correlations, and 
calculate the $^{17}$O NMR line broadening induced by
impurity doping.
\end{abstract}
\pacs{}]

%************************************************************
%***** Main Text*********************************************
%************************************************************

The most striking features observed in the normal state 
of underdoped cuprates are the occurance of a magnetic 
pseudogap and the persistence of antiferromagnetic (AF) correlations 
in the metallic state. Experimentally, the local magnetic properties of 
the cuprates can be sensitively probed by introducing impurities into 
the magnetically active Cu sites of the CuO$_2$ planes and by subsequently 
measuring the effect on the NMR signal of nuclei coupled to the planes.

Magnetic Ni ($d^8$) and nonmagnetic Zn ($d^{10}$) substituting for
Cu ($d^9$) are both found to introduce magnetic moments into the planes. 
Measurements of the macroscopic
susceptibility show an almost perfect $1/T$ Curie behavior,
the moments being stronger for Ni than for Zn~\cite{MEN94}.
Recently, Bobroff et al.~\cite{BOB97} presented novel $^{17}$O
NMR measurements for YBa$_2$(Cu$_{1-x}$(Zn/Ni)$_x$)$_3$O$_{6.6}$
probing the long-range polarization of the CuO$_2$ planes.
In contrast to the aforementioned bulk experiment, Zn had
a larger impact on the NMR line width than Ni, indicating a very 
different spatial variation of the polarization induced by the two
types of impurities. 
Furthermore, a markable deviation from the expected $1/T$ behavior of 
the line broadening 
was observed, which was interpreted by Morr et al.~\cite{SLI98} as an
indication of a temperature dependence of the AF correlation length $\xi$.
Here we present a microscopic theory of the impurity-induced 
local spin polarization in underdoped cuprates. We show that the 
presence of the spin gap and of short-range AF correlations strongly 
modifies the conventional RKKY picture which explains the
peculiarities in NMR line broadening data.

The planar Cu spins are described by a Heisenberg model treated in 
resonance valence bond (RVB) mean-field theory
\begin{equation}
H_{\text{RVB}} = -\sum_{\langle ij \rangle,\sigma} \left(
\Delta_{ij} f^{\dagger}_{i\sigma} f_{j\sigma}+\text{H.c.}\right).
\label{RVB}
\end{equation}
Dividing the lattice into two sublattices,
the symmetry of the order parameter $\Delta_{ij}$ is chosen
such as to obtain a pseudogap in the magnetic excitation 
spectrum~\cite{AZK88}
with density of states $\rho(\omega) = |\omega|/D^2$, where
$D$ is the spinon bandwidth; implicitly we assume this magnetic gap to 
be stabilized by the motion of holes present in the doped system. 
Magnetic and nonmagnetic impurities at site 0 are modelled by
\begin{eqnarray}
H_{\text{Ni}} &=& H'_{\text{RVB}} + J'\sum_{\delta} \bbox{s}_{\delta} 
\cdot \bbox{S}_0 ,\\
H_{\text{Zn}} &=& H_{\text{RVB}} + \lambda 
f^{\dagger}_0 f_0|_{\lambda\rightarrow \infty},
\end{eqnarray}
respectively, where $\bbox{S}_0$ is the spin of Ni and $\bbox{s}_{\delta}$
that of its nearest neighbor Cu sites, and $H'_{\text{RVB}}$ is obtained 
from Eq.~(\ref{RVB}) by excluding site 0 from the sum over bonds. The 
$S=1$ Ni spin is represented by two ferromagnetically coupled 
$S=1/2$ spins with coupling constant $|J_c|\gg|J'|$.

In the spin gap phase, Ni and Zn both induce moments on the 
Cu sites in the vicinity of the impurity~\cite{KHA97}. 
In the case of Ni, these moments themselves partially
screen the inherent spin of Ni, resulting in an effective $S=1/2$ moment
localized on the Ni site. The situation resembles that of an
underscreened Kondo problem; the Kondo energy 
$\omega_k \propto J'/\ln J'$ is, however, found to exhibit a power-law 
rather than the 
usual exponential dependence on the coupling parameter $J'$ stemming
from the fact that the Ni spin couples predominantly to the spinon bound state 
but not to extended states.
The formation of the Kondo state renormalizes the magnetic gap
but does not fill it.
The low-energy fixed point of the system is therefore that of an external 
$S=1/2$ spin coupled ferromagnetically to an impurity-free spin 
lattice with renormalized gap. 
The impurity-induced long-range polarization of the planar spins is
$K_{\text{Ni}}(\bbox{R},T) \propto R^{-3} T^{-1}$ for $\bbox{R}$ pointing
to the sublattice not containing the impurity, 
while it is found to be negligible otherwise. 
We notice that the rapid ($\propto R^{-3}$) decay, which should
be compared to the conventional Ruderman-Kittel polarization in
two dimensions ($\propto R^{-2})$, is due to the presence of the 
spin pseudogap, a feature which is robust against Ni doping.

In the case of Zn 
which carries no internal spin, the moments induced in the neighborhood 
of the impurity are not screened.
The moment associated with Zn is therefore only weakly 
localized, resulting in a slower spatial 
decay of the polarizability $K_{\text{Zn}}(\bbox{R},T) 
\propto R^{-2} (T\ln(D/T))^{-1}$ as compared to Ni; for $\bbox{R}$ pointing 
to the impurity sublattice, $K_{\text{Zn}}(\bbox{R},T)$ is negligible.

The broadening $\Delta \nu$ of the NMR line of a given planar
$^{17}O$ is dominated by the polarization of the two neighboring Cu 
sites. Since 
$K(\bbox{R},T)$ decays more slowly with distance for Zn than for Ni one 
expects $\Delta \nu_{\text{Zn}} > \Delta \nu_{\text{Ni}}$ for small
impurity concentration which is in agreement with experiment. 
However, the non-Curie behavior of $\Delta \nu$ observed by Bobroff et al.\
is not yet captured in the expressions for $K(\bbox{R},T)$. We therefore
extend our 
theory to include short-range AF correlations underestimated in the above 
mean-field treatment. Within a random-phase approximation (RPA) one finds
$K_{\text{RPA}}(\bbox{q},T) = K(\bbox{q},T) S_{\bbox{q}}(T)$
with the Stoner enhancement factor
\begin{equation}
S_{\bbox{q}}(T) =
\frac{\xi^2(T)}{1+(\bbox{q}-\bbox{Q})^2\xi^2(T)}
\end{equation}
for momenta $\bbox{q}$ close to $\bbox{Q} = (\pi,\pi)$.
At large enough distance the polarization induced by
Ni or Zn, respectively, is
\begin{eqnarray}
K_{\text{Ni}}(\bbox{R},T) &=& \pm \frac{3}{2\pi^3}
\frac{1}{R^3}\frac{\xi^2(T)}{T},
\label{KNI}\\
K_{\text{Zn}}(\bbox{R},T) &=& \pm \frac{1}{8\pi} 
\frac{1}{R^2}\frac{\xi^2(T)}{T\ln(D/T)},
\label{KZN}
\end{eqnarray}
where the alternating sign refers to the two different sublattices.
The NMR frequency shift induced by a given impurity
at position $\bbox{R}$ relative to a $^{17}O$ nucleus is
\begin{equation}
\Delta \omega = \gamma_e A_{\text{hf}} H 
\left[ K(\bbox{R}-\bbox{\delta}/2,T) + K(\bbox{R}+\bbox{\delta}/2,T) \right] ,
\label{FSH}
\end{equation}
where $A_{\text{hf}}$ is the n.n.\ hyperfine coupling constant, 
$\gamma_e$ the electronic magnetogyric ratio, $H$ the external magnetic 
field, and $\bbox{\delta}$ a lattice vector.
Due to the alternating sign of the polarizability and the 
fact that each $^{17}O$ nucleus lies in between two different
sublattices, the term in brackets in 
Eq.~(\ref{FSH}) partially vanishes; thus effectively the spatial derivative 
of $K(\bbox{R},T)$ is probed by NMR.
The  impurity-induced line broadening is finally calculated from
Eqs.~(\ref{KNI}), (\ref{KZN}), and (\ref{FSH})  
employing the formalism given in~\cite{WAL74}
to average over random impurity configurations, yielding
\begin{eqnarray}
\Delta \nu_{\text{Ni}} &\approx & 1.32 \gamma_e A_{\text{hf}} H x^{2}
\frac{\xi^2(T)}{T},\\
\Delta \nu_{\text{Zn}} &\approx & 0.77 \gamma_e A_{\text{hf}} H x^{3/2}
\frac{\xi^2(T)}{T\ln(D/T)}
\end{eqnarray}
with impurity concentration $x$.

\begin{figure}
\noindent
\centering
\epsfxsize=0.9\linewidth
\epsffile{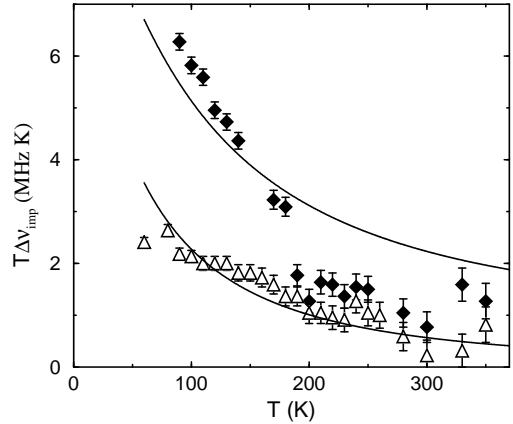}\\
\caption{Impurity-induced line broadening $T\Delta \nu$. 
The theoretical result is indicated by solid lines fitted to
$^{17}$O NMR data for 
$1\%$ Ni-doped (triangles) and $1\%$ Zn-doped (diamonds) 
YBa$_2$(Cu$_{1-x}$(Zn/Ni)$_x$)$_3$O$_{6.6}$~\protect\cite{BOB97}.}
\label{FIG1}
\end{figure}

We assume a phenomenological form 
$\xi(T) = 1/(a+bT)$ for the AF correlation length~\cite{STO97}, 
where $a$ and $b$ are fitting parameters of our theory.
In Fig.~\ref{FIG1} the result for $T\Delta \nu$ is
compared to experimental $^{17}$O NMR data. We choose
$a=0.07$ and $b=0.0007$ corresponding to $\xi=4.8$ in units of 
lattice spacings at $T=200 \text{~K}$, which compares to $\xi=5.9$
obtained in~\cite{BAR95}. We note that saturation of the AF correlation 
length which is expected to occur at low temperatures is not yet
included in the phenomenological expression used for $\xi(T)$.

In conclusion, we have presented a theory of magnetic and
nonmagnetic impurities in the normal state of underdoped cuprates. 
Based on the existence of a spin gap and extended to account for
the presence of AF correlations the theory gives a description
of the different magnetic behavior of Ni and Zn impurities and explains
well recent experimental results for the 
impurity-induced $^{17}$O NMR line broadening.

\end{document}